\documentclass[aps,pra,twocolumn,groupedaddress,amsmath,amssymb]{revtex4}

\usepackage{graphicx}

\begin{document}

\title{Stability and Decay Rates of Non-Isotropic Attractive Bose-Einstein
Condensates}

\author{
C. Huepe$^{1,4}$, L.S. Tuckerman$^{2}$, S. M{\'e}tens$^{3,4}$,
and M. E. Brachet$^{4}$\\
$^{1}$ {\it Department of Engineering Sciences and Applied Mathematics,
Northwestern University,
2145 Sheridan Road, Evanston, IL 60208, USA} \\
$^{2}$ {\it Laboratoire d'Informatique pour la M\'ecanique et les Sciences
de
l'Ing\'enieur, BP 133, 91403 Orsay, France} \\
$^{3}$ {\it Laboratoire de Physique Th\'eorique de la Mati\`ere
Condens\'ee,
Universit\'e Paris VII, Paris, France}\\
$^{4}$ {\it Laboratoire de Physique Statistique de l'Ecole Normale
Sup{\'e}rieure, \\
associ{\'e} au CNRS et aux Universit{\'e}s Paris VI et VII,
24 Rue Lhomond, 75231 Paris, France}
}

\date{\today}

\begin{abstract}
Non-Isotropic Attractive Bose-Einstein condensates are investigated with
Newton and inverse Arnoldi methods.  The stationary solutions of the
Gross-Pitaevskii equation and their linear stability are computed.
Bifurcation diagrams are calculated and used to find the condensate decay 
rates corresponding
to macroscopic quantum tunneling, two-three body inelastic collisions and
thermally induced collapse.

Isotropic and non-isotropic condensates are compared.
The effect of anisotropy on the bifurcation diagram and the decay rates is 
discussed.
Spontaneous isotropization of the condensates is found to occur. The 
influence of isotropization
on the decay rates is characterized near the critical point.
\end{abstract}

\maketitle

\tableofcontents

\section{Introduction}
\label{Introduction}
Experimental Bose-Einstein condensation
(BEC) with attractive interactions was first realized in ultra cold vapors
of {$~^7$Li} atoms \cite{Bradley1}, opening a new field in the study of
macroscopic quantum phenomena.
Such attractive condensates are known
to be metastable in spatially localized systems, provided that the number
of condensed particles is below a critical value ${\cal{N}}_c$
\cite{Bradley2}.
More recently,
Feshbach resonances in BEC of $~^{85}$Rb atoms were used
to investigate the stability and dynamics
of condensates with two-body interactions going from repulsive to
attractive values \cite{PRLRoberts}.

Experimental atomic traps generally use 
a harmonic and slightly asymmetric potential.
Thus, for most of the condensates produced so far, the geometry is nearly
spherical.
However, extremely asymmetric traps have been recently employed in
experimental investigations of cigar-like \cite{expcig1,expcig2,expcig3}
or pancake-like \cite{expcrepe} condensates.

Various physical processes compete to determine the
lifetime of attractive condensates.
The processes considered in this paper are
macroscopic quantum tunneling (MQT) \cite{Stoof1,Ueda},
inelastic two and three body collisions (ICO) \cite{Shi,Dodd,Kagan} and
thermally induced collapse (TIC) \cite{Stoof1,Sackett}.
The MQT and TIC contributions have been evaluated in the literature
using a variational Gaussian approximation to the condensate wave function.
However,
this approximation is known to be in substantial quantitative error --
e.g. as high as $17\%$ for ${\cal{N}}_c$ \cite{Ueda,Ruprecht,PRLiso} --
when compared to the exact solution of the Gross-Pitaevskii (G-P) equation.

In the nearly spherical isotropic case, both the elliptic (stable) 
and the hyperbolic (unstable) exact 
stationary solutions of the G-P equation were obtained 
numerically by Newton's method in \cite{PRLiso}. 
These solution branches were shown to meet at ${\cal{N}}_c$ 
through a generic Hamiltonian Saddle Node (HSN) bifurcation. 
While the Gaussian approximation presents an analogous HSN bifurcation, 
the amplitudes of its associated scaling laws were found to be in 
substantial ($\geq 14\%$) error. 
A method for computing the unstable branch in the isotropic case 
via a shooting method was outlined in \cite{Gammal1999}, but 
generalizing this procedure to higher dimensions would 
be inefficient, and impossible in non-rectangular domains.
The decay rates for the processes of MQT, ICO and TIC
were also computed, in the spherical case, from the numerical G-P
solutions.
They were shown to obey universal scaling laws.
Experimentally significant quantitative differences
were found between the exact rates and those based on the Gaussian
approximation \cite{PRLiso}.

In the extreme anisotropic cases, the variational Gaussian approximation
has been computed and compared to the G-P solution
on the elliptic (stable) branch \cite{GammalPRA64,GammalPreprint}.
This has allowed a more reliable determination of the
critical value ${\cal{N}}_c$ than can be obtained 
by the Gaussian approximation \cite{GammalPreprint}.
However, a faithful
determination of the lifetimes needs the computation of the hyperbolic
(unstable) branch \cite{PRLiso},
which has not yet been performed in the anisotropic case.

The main purpose of the present paper is to show that it is possible to
compute the full
HSN bifurcation diagram, and the corresponding lifetimes, in extreme
anisotropic cases.
We will do so by studying a cigar-like and a pancake-like condensate,
and will obtain their MQT, ICO and TIC decay rates.
While we have concentrated, for simplicity, on these two axisymmetric
cases,
the new numerical methods developed in this work are capable of solving
the general anisotropic problem.

The paper is organized as follows.
In section
\ref{Presentation of the Model}
we present the model considered throughout this work. After defining our
working
form of the G-P equation, we explain the methods that we used
to obtain the
stationary states and their linearized stability.
Section
\ref{Bifurcation}
is devoted to the numerical determination of the bifurcation diagram
and stability of the stationary states.
Isotropic and non-isotropic cases are compared and the dynamics is
discussed
in terms of the HSN bifurcation.
In section
\ref{Lifetime of Condensates}
we define and compute the decay rates.
Isotropic and non-isotropic rates are discussed,
their similarity is analyzed in terms of the
spontaneous isotropization of condensates.
Finally, section
\ref{Conclusion} is our conclusion.
Details of our numerical methods are given in the Appendix.


\section{Presentation of the Model}
\label{Presentation of the Model}

\subsection{Gross-Pitaevskii Equation}
\label{Gross-Pitaevskii Equation}

At low enough temperatures, neglecting the thermal and quantum
fluctuations, a Bose condensate can be represented by a complex wave
function $\Psi({\bf x},t)$ that obeys the dynamics of the G-P equation
\cite{Gross0,Pitaevskii1}.
Specifically, we consider a condensate of
${\cal{N}}$ particles of mass $m$ and (negative) effective scattering
length $\tilde{a}$ in a confining harmonic potential
$V(\tilde{\bf x}) =
m (\tilde{\omega}_x^2 \tilde{x}^2+\tilde{\omega}_y^2 \tilde{y}^2 +
\tilde{\omega}_z^2 \tilde{z}^2)/2$
where $\tilde{\bf x}=(\tilde{x},\tilde{y},\tilde{z})$
is the position vector.

These variables can be rescaled with respect to any
reference frequency $\hat{\omega}$ by using the natural quantum
harmonic oscillator units of time $\tau_0 = 1/\hat{\omega}$ and length
$L_0 = \sqrt{ \hbar / m \hat{\omega} }$.
In terms of the non-dimensional variables $t = \tilde{t}/\tau_0$,
${\bf x} = {\bf \tilde{x}}/L_0$, $a = 4 \pi \tilde{a}/ L_0$,
$\omega_x = \tilde{\omega}_x/ \hat{\omega}$,
$\omega_y = \tilde{\omega}_y/ \hat{\omega}$ and
$\omega_z = \tilde{\omega}_z / \hat{\omega}$,
the condensate is described by the action
\begin{equation}
\label{action}
{\cal{A}} = \int dt \left\{
\int d^3 x \frac{i}{2}
\left( \bar{\Psi} \frac{\partial \Psi}{\partial t} -
\Psi \frac{\partial \bar{\Psi}}{\partial t} \right)
- {\cal{F}} \right\},
\end{equation}
with
\begin{equation}
\label{Lagmult}
{\cal{F}} = {\cal{E}} - \mu {\cal{N}}
\end{equation}
where
\begin{eqnarray}
{\cal{N}} &=& \int d^3 x|\Psi|^2 \\
{\cal{E}} &=& \int d^3 x
\left[ \frac{1}{2} |\nabla \Psi|^2 +
V({\bf x}) |\Psi|^2 + \frac{a}{2} |\Psi|^4 \right] \\
V({\bf x}) &=& \frac{1}{2}(\omega_x^2 x^2 + \omega_y^2 y^2 + \omega_z^2 z^2
) \label{Vdef}
\end{eqnarray}

The Euler-Lagrange equation corresponding to ${\cal{A}}$ is
our working form of the Gross-Pitaevskii equation:
\begin{eqnarray}
\label{NLS}
- i \frac{\partial \Psi}{\partial t} &=&
- \frac{\delta \cal{F}}{\delta \bar{\Psi}} \\
&=& \left[\frac{1}{2} \nabla^2 - V({\bf x})
-\left( a |\Psi|^2 - \mu \right) \right] \Psi.
\nonumber
\end{eqnarray}

Our goal is to numerically determine the stable and unstable stationary
states of
(\ref{NLS}) and the eigenvalues of (\ref{NLS}) linearized
about these stationary states.
We will carry out this calculation for various values of
a cylindrical potential defined by
$\omega_r \equiv \omega_x = \omega_y$ and $\omega_z$: the isotropic
case $\omega_r= \omega_z$, a cigar case $\omega_r /5 = \omega_z$
and a pancake case $\omega_r = \omega_z / 5$.
We will then use these results to calculate the condensate
decay rates and compare these decay rates to those produced
by the Gaussian approximation.

\subsection{Stationary States}
\label{Stationary}
Stationary states of (\ref{NLS}) corresponding
to minima of ${\cal{E}}$ at a given value of ${\cal{N}}$
can be obtained by integrating to relaxation the diffusion equation
\begin{eqnarray}
\label{GLRmeb}
\frac{\partial \Psi}{\partial t} &=&
-\frac{\delta \cal{F}}{\delta \bar{\Psi}} =
\left[\frac{1}{2} \nabla^2 - V({\bf x})
-\left( a |\Psi|^2 - \mu \right) \right] \Psi \\
\label{conserve}
\frac{\partial {\cal{N}}}{\partial t} &=& 0,
\end{eqnarray}
using initial data $\Psi(t=0)$ with a total number of particles
${\cal{N}}$.
The condition (\ref{conserve}) fixes
the value of the Lagrange multiplier $\mu$
during the relaxation.
This relaxation method yields both the solution $\Psi$ and
the Lagrange multiplier $\mu$. It is equivalent to
that used in \cite{Ruprecht} and
can only reach the stable stationary solutions of (\ref{GLRmeb}).
Unstable stationary solutions to (\ref{NLS}) and (\ref{GLRmeb})
are obtained by a Newton branch following method detailed in the Appendix.

\begin{figure}
\includegraphics[width=8cm,height=7cm]{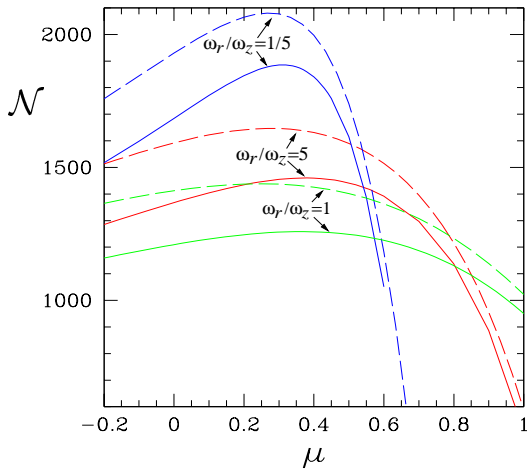}
\caption{\label{fig:N_vs_mu}
Particle number ${\cal{N}}$ as a function of $\mu$ for
the exact solutions (solid curves) and the
Gaussian approximation (dashed curves) presented in section
\ref{Bifurcation}.
From top to bottom: pancake ($\omega_r=\omega_z/5$), 
cigar ($\omega_r/5=\omega_z$), and isotropic ($\omega_r=\omega_z$) 
geometries.}
\end{figure}

Note that the Lagrange multiplier $\mu$ can only affect the solutions of
(\ref{NLS})
through a homogeneous rotating phase factor $e^{i \mu t}$,
in contrast to its particle number conservation effect on equation
(\ref{GLRmeb}).
However, every stationary solution to (\ref{NLS}) is indexed by the unique
$\mu$-value
that makes it time-independent, as shown in figure \ref{fig:N_vs_mu}.

\subsection{Linearized Stability}
\label{Linearized Stability}

We now turn our attention to computing the linear stability
of the G-P equation about a stationary state.
We first write (\ref{NLS}) in the abbreviated form:
\begin{equation}
-i\frac{\partial \Psi}{\partial t} = L \Psi + W(\Psi)
\label{abbrev}
\end{equation}
where
\begin{eqnarray}
L \Psi &\equiv& \frac{1}{2}\nabla^2 \Psi \label{Ldef}\\
W(\Psi) &\equiv& \left[-V({\bf x})
- a |\Psi|^2 + \mu \right] \Psi \label{Wdef}
\end{eqnarray}
The stationary states of (\ref{abbrev}) satisfy:
\begin{equation}
0=L\Psi + W(\Psi)
\label{steady}\end{equation}

Without loss of generality, $\Psi$ can be chosen to be real. 
Our objective is to calculate 
the eigenpairs $(\lambda,\psi)$ of the operator that results 
from linearizing (\ref{abbrev}) about a stationary state $\Psi$. 
(We use $\Psi$ to designate solutions to the nonlinear 
problem (\ref{steady}) and $\psi$ to designate eigenvectors, 
which are solutions to the linear problem to be defined below.) 
In order to correctly formulate the linear stability problem, 
it is necessary to first decompose $\psi = \psi^R + i \psi^I$.
We write the linearized evolution equation
\begin{equation}
\frac{\partial\psi}{\partial t} = i(L+DW(\Psi))(\psi^R+ i \psi^I)
\label{linstab}\end{equation}
where $DW(\Psi)$ is the Fr\'echet derivative, or Jacobian, of $W$
evaluated at $\Psi$. $DW(\Psi)$ acts on $\psi$ via:
\begin{equation}
DW \psi = DW^R \psi^R + i DW^I \psi^I
\end{equation}
where we have omitted the functional dependence of
$DW$, $DW^R$, and $DW^I$ on $\Psi$, and where
\begin{subequations}
\begin{eqnarray}
DW^R &\equiv& \mu-V({\bf x}) -3a\Psi^2 \\
DW^I &\equiv& \mu-V({\bf x}) -a\Psi^2.
\end{eqnarray}
\end{subequations}
Equation (\ref{linstab}) is then written in matrix form as
\begin{equation}
\frac{\partial}{\partial t}\left(\begin{array}{c}
\psi^R \\ \psi^I \end{array} \right) =
\left[\begin{array}{c c} 0 & -(L+DW^I) \\ L+DW^R & 0 \end{array} \right]
\left(\begin{array}{c} \psi^R \\ \psi^I \end{array} \right)
\end{equation}
%
The eigenmodes $(\lambda,\psi^R,\psi^I)$ satisfy
\begin{equation}
\lambda\left(\begin{array}{c}
\psi^R \\ \psi^I \end{array} \right) =
\left[\begin{array}{c c} 0 & -(L+DW^I) \\ L+DW^R & 0 \end{array} \right]
\left(\begin{array}{c} \psi^R \\ \psi^I \end{array} \right) \label{matrix}
\end{equation}
Note that  this eigensystem is usually presented in the literature
(see, for example,
reference \cite{Griffin})
in terms of the variables
\begin{equation}
(\omega^{\rm B},\psi,\psi^*) \equiv 
(-i \lambda,\psi^R + i \psi^I ,\psi^R - i \psi^I)
\end{equation}
as the equivalent
Bogoliubov--de Gennes coupled equations
\begin{equation}
\omega^{\rm B}\left(\begin{array}{c}
\psi \\ \psi^* \end{array} \right) =
\left[\begin{array}{c c} L + DW^{\rm B}  & -a\Psi^2  \\
a\Psi^2  & -(L - DW^{\rm B})  \end{array} \right]
\left(\begin{array}{c} \psi \\ 
\psi^* \end{array} \right) \label{matrix_bog}
\end{equation}
where
\begin{equation}
DW^{\rm B} \equiv \mu-V({\bf x}) -2a\Psi^2
\end{equation}
In the following, we will work with
matrix formulation (\ref{matrix}) because it avoids a potential
notational inconsistency of  (\ref{matrix_bog}) arising
from the fact that $\psi$ and $\psi^*$ are complex conjugates only when 
$\omega^{\rm B}$ is imaginary.

It is more convenient to work with the square of the matrix in
(\ref{matrix}):
{\footnotesize
\begin{eqnarray}
&&
\lambda^2 \left(\begin{array}{c}
\psi^R \\ \psi^I \end{array} \right) = \label{sqrmatrix}\\
&&
\left[\begin{array}{c c} -(L+DW^I)(L+DW^R) & 0 \\
0 & -(L+DW^R)(L+DW^I) \end{array} \right]
\left(\begin{array}{c} \psi^R \\ \psi^I \end{array} \right) \nonumber
\end{eqnarray}
}
Because (\ref{sqrmatrix}) is block diagonal, it can be separated into
the two problems:
\begin{subequations}
\begin{eqnarray}
\lambda^2 \psi^R = -(L+DW^I)(L+DW^R) \psi^R \label{proba}\\
\lambda^2 \psi^I = -(L+DW^R)(L+DW^I) \psi^I \label{probb}
\end{eqnarray}
\end{subequations}
Problems (\ref{proba}) and (\ref{probb}) are closely related,
Since the operators $L$, $DW^I$, and $DW^R$ are all self-adjoint
under the standard Euclidean inner product,
the operators in (\ref{proba}) and (\ref{probb}) are adjoint to each other.
If $\psi^R$ is an eigenvector of (\ref{proba}) with
non-zero eigenvalue $\lambda^2$, then $(L+DW^R)\psi^R$ is an eigenvector
of (\ref{probb}) with the same eigenvalue.
(Similarly, if $(\lambda,\psi^R,\psi^I)$ is an eigenmode of (\ref{matrix}),
then $(-\lambda,\psi^R,-\psi^I)$ is also an eigenmode of (\ref{matrix}).)
Thus, we solve only (\ref{proba}).
The eigenvalues $\lambda^2$ of (\ref{proba})-(\ref{probb})
must be either complex conjugate pairs or real.
We find them to be real and (almost all) negative,
perturbed only slightly from the eigenvalues of $-L^2$.
The eigenvalues $\lambda$ of (\ref{matrix}) are therefore
found to be either pure imaginary or pure real, with most imaginary.

Problems (\ref{matrix}) and (\ref{proba})-(\ref{probb}) have
neutral eigenmodes which reflect the physical invariances
of the problem.
Since $DW^I\Psi = W(\Psi)$, then the stationary state
$\Psi$ is a neutral mode of $L+DW^I$ and hence of problem (\ref{probb}).
This neutral mode is the phase mode of (\ref{NLS}),
since its existence is a consequence of the invariance of solutions
$\Psi$ to (\ref{steady}) under multiplication by any complex number
on the unit circle.
The corresponding eigenmode of problem (\ref{proba}) is
$d \Psi / d \mu$. This neutral mode can be understood as
a consequence of differentiating (\ref{NLS}) with respect to $\mu$:
\begin{equation}
0 = \frac{d}{d\mu}[(L+W)\Psi] = (L+DW^R)\frac{d\Psi}{d\mu} + \Psi
\label{muderiv}
\end{equation}
Thus,
\begin{equation}
-(L+DW^I)(L+DW^R)\frac{d\Psi}{d\mu} = (L+DW^I)\Psi = 0.
\end{equation}
In terms of the original problem (\ref{matrix}), the phase mode
$(\lambda,\psi^R,\psi^I)=(0,0,\Psi)$ is a neutral eigenvector, while
(\ref{muderiv}) shows that
$(\lambda,\psi^R,\psi^I)=(0,d\Psi/d\mu,0)$
is a neutral generalized eigenvector,
the two modes forming a Jordan block for (\ref{matrix}).

In practice, we fix $\mu$ to calculate the stationary
states $\Psi$ and the eigenvalues.
The operators of (\ref{proba})-(\ref{probb}) depend on $\mu$
both explicitly and through $\Psi$.
For $\mu$ above a critical value $\mu_c$ all eigenvalues
$\lambda$ are imaginary, i.e. $\Psi$ is an elliptic stationary
state of (\ref{abbrev}).
As $\mu$ crosses $\mu_c$, we will see that one imaginary pair
fuses at zero, and becomes real, with one positive and one
negative value of $\lambda$ for $\mu< \mu_c$.
Stationary states for $\mu<\mu_c$ are thus
hyperbolic in the directions corresponding to these eigenvalues.

\section{Bifurcation and Stability of Condensates}
\label{Bifurcation}

In this section we will find the stationary solutions and
study the stability of isotropic ($\omega_r=\omega_z$),
cigar-like ($\omega_r /5 = \omega_z$) and pancake-like
($\omega_r = \omega_z / 5$) condensates.
These results were obtained by solving equation (\ref{steady}) for
the stationary states and (\ref{matrix}) or (\ref{proba}) for the
corresponding bifurcating eigenvalues.
The system is discretized using pseudo-spectral methods
in a spherical domain for the isotropic case and in a
periodic Cartesian domain for the non-isotropic cases.
We use Newton's method to calculate the branches of stationary states.
The bifurcating eigenvalue is found in the
isotropic case by diagonalizing the
matrix corresponding to (\ref{matrix}).
In the non-isotropic case,
we use instead the iterative inverse Arnoldi method, which
requires only actions of the operator in (\ref{proba}).
The BiCGSTAB variant of the conjugate gradient
method is used to solve the linear systems required by both
Newton's method and the inverse Arnoldi method.
The numerical methods we use are described in greater detail in the Appendix.

\subsection{Isotropic Condensate}
\label{Isotropic Condensate}

\begin{figure}
\includegraphics[width=8cm,height=7cm]{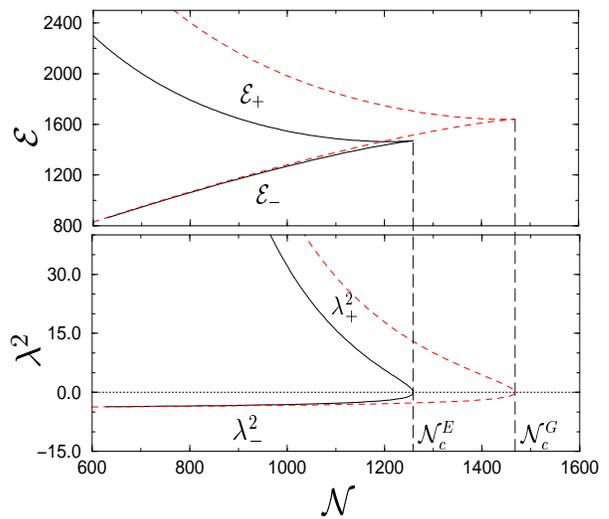}
\caption{\label{fig:isotropic_eigs}
Stationary solutions of the GP equation versus the
particle number ${\mathcal{N}}$ for the isotropic potential
case with
$\hat{\omega}_r = \hat{\omega}_z = \hat{\omega}$.
Top: value of the energy functional ${\mathcal{E}}_+$ on the 
unstable (hyperbolic) branch and ${\mathcal{E}}_-$ on the 
stable (elliptic) branch.
Bottom: square of the bifurcating eigenvalue
($\lambda_{\pm}^2$). Note that $|\lambda_-|$ is the energy of small
excitations around the stable branch.
Solid lines: exact solution of the GP equation.
Dashed lines: Gaussian approximation.
}
\end{figure}
\begin{figure}
\includegraphics[width=7cm]{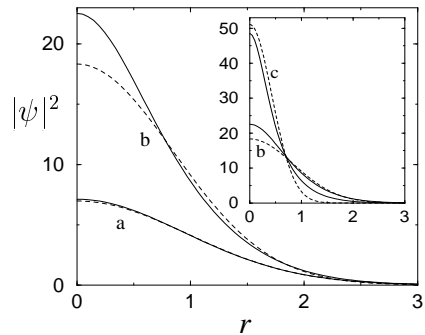}
\caption{\label{fig:psi}
Condensate density $|\psi|^2$ as a function of radius $r$, in
reduced units (see text). Solid lines: exact solution
of the G-P equation.
Dashed lines: Gaussian approximation.
Stable (elliptic) solutions are shown for particle number
${\cal{N}}=252$ (a) and
${\cal{N}}=1132$ (b).
(c) is the unstable (hyperbolic) solution for
${\cal{N}}=1132$ (see insert).
}
\end{figure}

In order to compare our results with the existing experiments on
quasi-isotropic condensates, we will use the following physical
constants, corresponding to {$^7$Li} atoms in a radial trap:
$m = 1.16 \times 10^{-26} \, {\mathrm{kg}}$,
$\tilde a = -27.3 a_0$ (with $a_0 $ the Bohr radius) and
$\hat \omega= (\tilde{\omega}_x \tilde{\omega}_y
\tilde{\omega_z})^{1/3}=908.41 \,{\mathrm{s^{-1}}}$.
These values yield $a = -5.74 \times 10^{-3}$.
With these parameters,
the mean-field approximation (\ref{NLS}) is
expected to be very reliable.
Note that we ignore the contributions of non-condensed atoms.
They interact with the condensate only
through a nearly constant background density term,
inducing no significant change in the dynamics of the
system \cite{Hourbiers}.

The values of the energy functional
${\cal{E}}$ and the (smallest absolute value) square
eigenvalue $\lambda^2$ versus particle number
${\cal{N}}$ are shown as solid lines
on figure \ref{fig:isotropic_eigs} (top and bottom, respectively).
The eigenvalues are imaginary on the metastable elliptic lower branch
($\lambda^2 < 0$) and real on the unstable hyperbolic upper branch
($\lambda^2 > 0$).
Using (\ref{Lagmult})
on stationary solutions we obtain $d{\cal{E}}/d{\cal{N}}=\mu$.
Thus $\mu $ is the slope of ${\cal{E}}$ and
the lower branches ${\cal{E}}_{-}$, $\lambda^2_{-}$ (respectively
upper branches ${\cal{E}}_{+}$, $\lambda^2_{+}$)
are scanned for $\mu > \mu _c$ (respectively $\mu < \mu _c$).
The point $\mu = \mu _c$ determines the maximum number of particles
${\cal{N}} = {\cal{N}}_c$ for which stationary solutions exist.
We have checked
that all the other pairs of eigenvalues are
imaginary on both branches (data not shown).

The dashed curves on figure \ref{fig:isotropic_eigs}
are derived from the Gaussian variational approximation which will be
defined in section \ref{Non-isotropic Condensates} for the general
anisotropic case.
In the present isotropic case, this approximation
can be solved in closed form, yielding the expressions \cite{PRLiso}:
\begin{eqnarray}
\label{eqNgauss}
{\cal{N}}(\mu ) =
\frac{4 \sqrt{2 \pi^3} \left(-8 \mu + 3 \sqrt{7 + 4 \mu ^2} \right)}
{7 |\tilde{a}| \left(-2 \mu + \sqrt{7 + 4 \mu ^2} \right)
^{3/2}}, \\
\label{eqEgauss}
{\cal{E}} = {\cal{N}}(\mu ) \left(-\mu + 3 \sqrt{7 + 4 \mu ^2} \right) / 7.
\end{eqnarray}
The number of particles ${\cal{N}}$ is maximal at
${\cal{N}}^G_c = 8 \sqrt{2 \pi^3} / |5^{5/4} a|$
for $\mu = \mu ^G_c = 1 / 2 \sqrt{5}$.
The eigenvalues can also be obtained in closed form from the linearized
equations of motion \cite{PRLiso}:
\begin{equation}
\label{eqlambda}
\lambda^2(\mu ) = 8 \mu ^2 - 4 \mu \sqrt{7 + 4 \mu ^2} + 2
\end{equation}

By inspection of figure \ref{fig:isotropic_eigs} it is apparent that both
the
solution of the GP equation and the Gaussian variational approximation
share the same qualitative behavior, with quantitative discrepancies.
Figure \ref{fig:psi} shows the physical origin of the quantitative errors
in the Gaussian approximation. It is
apparent that the exact solution is well approximated by
a Gaussian only for small ${\cal{N}}$ on the stable (elliptic) branch.

\subsection{Hamiltonian Saddle Node Normal Form}
\label{Dynamics and Normal Form}

\begin{figure}
\includegraphics[width=9cm]{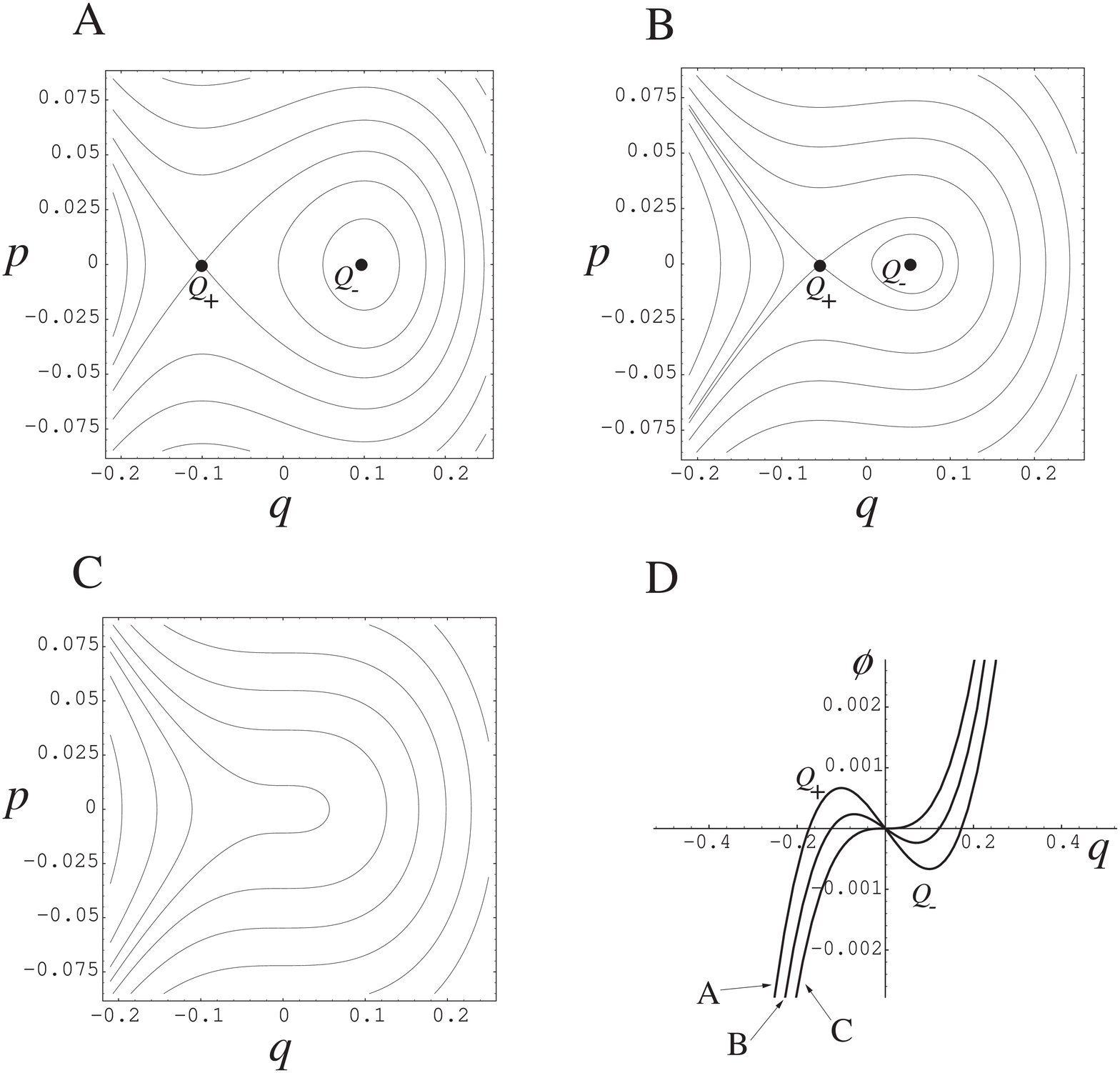}
\caption{\label{fig:PhsPrt}
Phase portraits of the Hamiltonian saddle-node
normal form (\ref{eqNF}), with
$p = \dot{q}$.
A: $\delta = 0.2$, B: $\delta = 0.1$, C: $\delta = 0$.
D: Corresponding potential $\Phi$ associated to each phase
portrait A, B or C, with $\dot p = -\partial \Phi / \partial q$.
An elliptic region bounded by the separatrix that starts and
ends on the fixed point $Q_+$ (homoclinic orbit) is present on A and B.
Phase portrait C displays the critical merging of fixed points
$Q_+$ and $Q_+$, and the disappearance of the elliptic region.
}
\end{figure}

The qualitative behavior displayed on figure
\ref{fig:isotropic_eigs} by the physical quantities
${\cal{E}}$ and $\lambda^2$ near the critical value
${\cal{N}}={\cal{N}}_c$ is the
generic signature of a HSN bifurcation
defined, at lowest order, by the normal form \cite{GH,Elphick}
\begin{equation}
\label{eqNF}
\ddot{q} = \delta - \beta q^2,
\end{equation}
where $\delta = (1-{\cal{N}}/{\cal{N}}_c)$ is the bifurcation
parameter, $\beta$ is a dimensionless constant, 
and $q$ is the coordinate describing the state of the system 
in the direction of the phase space that becomes unstable.
Indeed, introducing the additional non-dimensional quantities 
${\cal{E}}$ and $\gamma$ to define the appropriate energy
\begin{equation}
{\cal{E}} = {\cal{E}}_0 + \frac{1}{2} \dot{q}^2 - \delta q +
\frac{1}{3} \beta q^3 - \gamma \delta,
\end{equation}
it is straightforward to derive from (\ref{eqNF}) that,
close to the critical point $\delta=0$, the universal scaling laws
are given by
\begin{eqnarray}
\label{eqsclE}
{\cal{E}}_\pm &=&
{\cal{E}}_c - {\cal{E}}_l \delta \pm {\cal{E}}_{\Delta} \delta^{3/2}, \\
\label{eqscllam}
\lambda^{2}_\pm &=& \pm \lambda^{2}_{\Delta} \delta^{1/2},
\end{eqnarray}
where ${\cal{E}}_c={\cal{E}}_0$,
${\cal{E}}_l=\gamma$,
${\cal{E}}_{\Delta}=2 / 3 \sqrt{\beta}$ and
$\lambda^{2}_{\Delta}=2 \sqrt{\beta}$.
Note that these relations can be inverted to obtain the parameters in
(\ref{eqNF}) from the critical data. For the Gaussian approximation
the critical amplitudes can be computed from equations
(\ref{eqNgauss}) and (\ref{eqEgauss}).
One finds
\begin{eqnarray}
{\cal{E}}_c &=& \frac{4 \sqrt{2 \pi^3} }
{ 5^{3/4} |a|} \\
{\cal{E}}_{\Delta} &=& \frac{64 \sqrt{\pi^3} }
{ 5^{9/4} |a|} \\
\lambda^{2}_{\Delta} &=& 4 \sqrt{10}.
\end{eqnarray}
For the exact solutions, we obtain the critical amplitudes
${\cal{E}}_{\Delta} = 1340$ and $\lambda^{2}_{\Delta} = 14.68$
by performing fits on the numerical data.
Comparing both results, we find that the Gaussian approximation
captures the bifurcation qualitatively, but with quantitative
errors of $17\%$ for ${\cal{N}}_c$ \cite{Ruprecht},
$24\%$ for ${\cal{E}}_{\Delta}$ and
$14\%$ for $\lambda^{2}_{\Delta}$ 
in the isotropic case \cite{PRLiso}.

The phase portrait of the normal form is shown on Figure
\ref{fig:PhsPrt}.
When $\delta = (1-{\cal{N}}/{\cal{N}}_c)>0$, 
equation (\ref{eqNF}) admits
two fixed points $Q_\pm=\mp\sqrt{\delta/\beta}$,
as shown in Fig.~\ref{fig:PhsPrt}A.
Thus, a hyperbolic stationary state and an elliptic stationary state
coexist.
The phase space is separated into two regions by a separatrix which
is a homoclinic orbit
linking the hyperbolic stationary state to itself.
Trajectories inside the orbit remain bounded near the
elliptic fixed point. If the condensate is taken beyond the
separatrix by a perturbation
(e.g. thermal excitations or quantum tunneling,
see below section \ref{Lifetime of Condensates}), it will
fall into unbounded (hyperbolic) trajectories and collapse.
As ${\cal{N}}$ is increased, the hyperbolic and
elliptic stationary states approach one another,
(Fig.~\ref{fig:PhsPrt}B)
and the homoclinic orbit inside which orbits are bounded is reduced.
The two stationary states join at ${\cal{N}}={\cal{N}}_c$
(Fig.~\ref{fig:PhsPrt}C), at which the HSN occurs.
No stationary state exists for ${\cal{N}}>{\cal{N}}_c$.

\subsection{Non-isotropic Condensates}
\label{Non-isotropic Condensates}

\begin{figure}
\includegraphics[width=8cm,height=7cm]{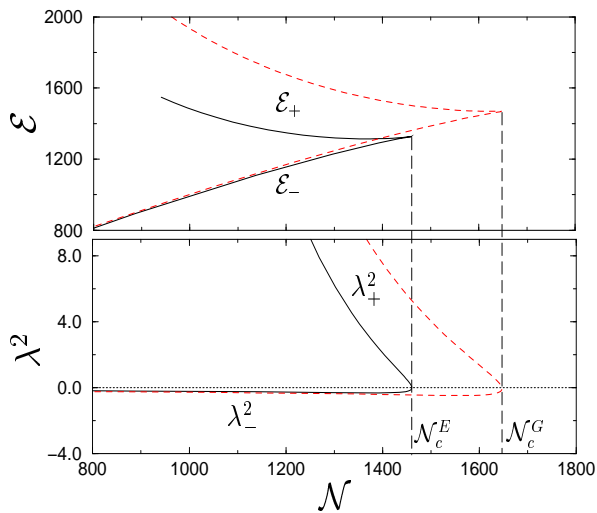}
\caption{\label{fig:cigar_eigs}
Stationary solutions of the GP equation versus the
particle number ${\mathcal{N}}$ for a non-isotropic potential
case with $\omega_r = \hat{\omega}$ and
$\omega_z = \hat{\omega}/5$ ({\emph{cigar}}).
Top: value of the energy functional.
Bottom: square of the bifurcating eigenvalue ($\lambda_{\pm}^2$).
Solid lines: exact solution of the GP equation.
Dashed lines: Gaussian approximation.
}
\end{figure}
\begin{figure}
\includegraphics[width=8cm,height=7cm]{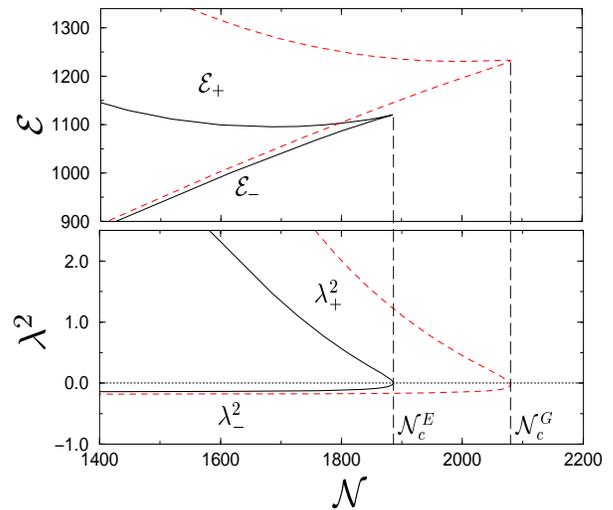}
\caption{\label{fig:pancake_eigs}
Stationary solutions of the GP equation versus the
particle number ${\mathcal{N}}$ for a non-isotropic potential
case with $\omega_r = \hat{\omega}/5$ and
$\omega_z = \hat{\omega}$ ({\emph{pancake}}).
Top: value of the energy functional.
Bottom: square of the bifurcating eigenvalue ($\lambda_{\pm}^2$).
Solid lines: exact solution of the GP equation.
Dashed lines: Gaussian approximation.
}
\end{figure}

We now briefly present the main expressions obtained from
a Gaussian variational analysis of the G-P equation
with a cylindrical potential trap. Some of these results have
been previously obtained by other authors
\cite{GammalPRA64,GammalPreprint,PerezGarciaAsym1,PerezGarciaAsym2,AbdullaevPRA63,AdhikariPRE65}.
We therefore restrict our discussion to the equations
that will be used in our analysis of the condensate lifetimes.

The trial function is a Gaussian solution to
the linear $(a=0)$ Schr\"{o}dinger equation in which
we incorporate eight variational parameters in order
to take into account the anisotropy of the system.
The form of the ansatz is given by:
\begin{eqnarray}\label{A1}
\Psi(x,y,z,t)=\left(A_r(t) + i A_i(t)\right)
\nonumber \\
\times \exp\left\{
-\left(\frac{1}{X(t)^2} + i \phi_X(t)\right) \frac{x^2}{2} \right .
\nonumber \\
-\left(\frac{1}{Y(t)^2} + i \phi_Y(t)\right) \frac{y^2}{2}
\nonumber \\ \left .
-\left(\frac{1}{Z(t)^2} + i \phi_Z(t)\right) \frac{z^2}{2}
\right\}
\end{eqnarray}
where the real parameters $\{A_r, A_i\}$, $\{\phi_X,\phi_Y,\phi_Z\}$ and
$\{X,Y,Z\}$ are related to the amplitude, the phase and the width of the
Gaussian profile respectively.
The Euler-Lagrange equations associated with the trial function (\ref{A1})
and the action
defined in equation (\ref{action})
can be reduced to the following system of
second-order
differential equations:

\begin{eqnarray}\label{A2}
\frac{d^2 X}{d t^2}&=& - \omega_x^2 X - \frac{\nu }{X^2 Y Z} +
\frac{1}{X^3}
\nonumber \\
\frac{d^2 Y}{d t^2}&=& - \omega_y^2 Y - \frac{\nu}{X Y^2 Z} +
\frac{1}{Y^3}
\nonumber \\
\frac{d^2 Z}{d t^2}&=& -\omega_z^2 Z - \frac{\nu }{Z^2 X Y } +
\frac{1}{Z^3}
\end{eqnarray}
where
\begin{equation}
\nu=\sqrt{\frac{2}{\pi}\frac{m\ (\omega_x \omega_y
\omega_z)^{1/3}}{\hbar}}
|\tilde{a}| {\cal N}.
\end{equation}
The evolution of the condensate
is better understood by drawing an analogy between its
width and the motion of a particle with coordinates
$(X,Y,Z)$ moving in the potential
\begin{eqnarray}\label{A2a}
U(X,Y,Z)&=& \frac{1}{2}
\left( \omega_x^2 X^2 + \omega_y^2 Y^2 + \omega_z^2 Z^2 \right) -
\frac{\nu}{ X Y Z} \nonumber \\ &+ & \frac{1}{2}
\left( \frac{1}{X^2} + \frac{1}{Y^2}+\frac{1}{Z^2} \right).
\end{eqnarray}
Indeed, defining $P_x = d X/d t$, $P_y = d Y/d t$, $P_z = d Z/d t$
and the Hamiltonian
\begin{eqnarray}\label{A2b}
& & H(P_x,P_y,P_z,X,Y,Z)= \frac{1}{2} \left( P_x^2 + P_y^2 +P_z^2 \right)
+ U(X,Y,Z), \nonumber
\end{eqnarray}
we find that equations (\ref{A2}) transform into 
Hamilton's equations of motion.

If we consider now a potential trap (\ref{Vdef}) with cylindrical
symmetry ($\omega_r \equiv \omega_x = \omega_y$) equations
(\ref{A2}) can be simplified by using $X(t)=Y(t)$.
We thus find that (\ref{A2}) yields two fixed points
$(X_+,Z_+)$ and $(X_-,Z_-)$ which describe the stationary solutions
for $\Psi(x,y,z,t)$.
These obey
\begin{eqnarray}\label{A3}
0&=& \omega_r^2 - \frac{4 \mu}{7 X_{\pm}^2} - \frac{5}{7 X_{\pm}^4}+
\frac{2}{7}\left( \frac{1}{X_{\pm}^2} + \frac{1}{Z_{\pm}^2} \right)
\frac{1}{X_{\pm}^2}\nonumber \\
0&=& \omega_z^2 - \frac{4 \mu}{7 Z_{\pm}^2} - \frac{5}{7 Z_{\pm}^4}+
\frac{4}{7}\left( \frac{1}{X_{\pm}^2 Z_{\pm}^2}\right),
\end{eqnarray}
where the chemical potential $\mu$ is related to the total number
of particles through
\begin{eqnarray}
\mathcal{N}&=& \frac{2 L_{0}}{7 |\tilde{a}|} \sqrt{2 \pi} X^2 Z \left(
\frac{1}{X^2} + \frac{1}{2 Z^2} - \mu \right).
\end{eqnarray}
The fixed points correspond
to a metastable center $(X_+,Z_+)$
and to an unstable saddle point $(X_-,Z_-)$, respectively.
They are analogous to the $Q_+$ and $Q_-$ points appearing in
the phase portraits on Figure \ref{fig:PhsPrt}.
The solutions to (\ref{A3}) can be computed numerically,
together with the linearized variational
equations evaluated at every stationary point.

Figures \ref{fig:cigar_eigs} and \ref{fig:pancake_eigs} show
${\cal{E}}$ and $\lambda^2$ for the cigar and pancake cases,
respectively. The solid lines present the values obtained by
discretizing and solving numerically the original differential
equations (\ref{steady}) and (\ref{proba}), using the methods
described in the Appendix.
The dashed lines were computed using the Gaussian approximation
described above.
Both the isotropic and non-isotropic cases display saddle-node
bifurcations.
This is to be expected, since the saddle-node bifurcation is
the generic way in which stable and unstable branches meet \cite{GH}.

It is apparent from figures \ref{fig:isotropic_eigs}, \ref{fig:cigar_eigs},
and \ref{fig:pancake_eigs} that the exact critical number of
particles ${\mathcal{N}}^E_c$ is smaller than the Gaussian value
${\mathcal{N}}^G_c$ for all three geometries
\cite{Ueda,Ruprecht,GammalPRA64,GammalPreprint}.
Table \ref{table_1} compares the different critical ${\mathcal{N}}$
values obtained.

\begin{table}
\caption{\label{table_1}
Critical number of particles obtained 
for the isotropic, cigar and pancake geometries by 
using the exact solution of the GP equation ($\mathcal{N}_c^E$)
and the Gaussian approximation ($\mathcal{N}_c^G$)}
\begin{ruledtabular}
\begin{tabular}{|cc|cc|}
\ $\omega_r$ \ & \ $\omega_z$ \ & \
${\mathcal{N}}^E_c$ \ & \ ${\mathcal{N}}^G_c$ \\
\hline \hline
\ $\hat{\omega}$ \ & \ $\hat{\omega}$
\ & \ $ 1258.5 $ \ & \ $ 1467.7 $ \ \\ \hline
\ $\hat{\omega}$ \ & \ $\hat{\omega} / 5$
\ & \ $ 1460.3 $ \ & \ $ 1646.6 $ \ \\ \hline
\ $\hat{\omega}/5$ \ & \ $\hat{\omega}$
\ & \ $ 1885.6 $ \ & \ $ 2080.5 $ \ \\
\end{tabular}
\end{ruledtabular}
\end{table}

In order to compare properly the HSN bifurcations obtained
for the three aspect ratios studied,
we can rescale the intensity of the potential to
obtain the same ${\mathcal{N}}_c^E$ for all cases.
In general, any confining harmonic potential with
frequencies $\omega_r$ and $\omega_z$ that produces
a critical number of particles ${\mathcal{N}}_c$
can be rescaled by a factor
\begin{equation} \label{def:cfactor}
c = \left( \frac{{\mathcal{N}}_c}{{{\mathcal{N}}_c}^*} \right)^2,
\end{equation}
to obtain the a new potential with
frequencies $\omega_r^* = c \, \omega_r$ and
$\omega_z^* = c \, \omega_z$, which will have the critical
number of particles ${\mathcal{N}}_c^*$.
The remaining physical quantities for the new potential are
obtained through the following transformations:
\begin{eqnarray}
\Psi^* &=& \frac{\Psi}{ c^{1/4} } \\
{\mathcal{N}}^* &=& \frac{{\mathcal{N}} }{ \sqrt{c} } \\
{\mathcal{E}}^* &=& \frac{{\mathcal{E}} }{ \sqrt{c} } \\
\lambda^* &=& \lambda.
\end{eqnarray}
We choose arbitrarily to rescale the potential intensity so that
all ${\mathcal{N}}_c^E$ are equal to that for the isotropic
case ${\mathcal{N}}_c^{\mathrm{iso}}$. Table \ref{table_2}
shows the value of the rescaling factors
$c_{\mathrm{cig}}$ and $c_{\mathrm{pan}}$ for the cigar and
pancake cases respectively, as obtained from equation
(\ref{def:cfactor}) using
${\mathcal{N}}_c^* = {\mathcal{N}}_c^{\mathrm{iso}}$.
The last two columns of this table show the critical amplitudes
obtained for the rescaled ${\mathcal{E}}$ and $\lambda^2$
curves. These were obtained by fitting the HSN asymptotic forms
given in relations (\ref{eqsclE}) and (\ref{eqscllam}) to the
rescaled data.

\begin{table}
\caption{\label{table_2} 
Rescaling factors required for having
$\mathcal{N}_c = \mathcal{N}_c^{\mathrm{iso}}$ in the
cigar and pancake cases. Columns ${\mathcal{E}_\Delta}$ and
$\lambda^2_\Delta$ show critical amplitudes at the bifurcation
for the rescaled ${\mathcal{E}}$ and $\lambda^2$ curves respectively.}
\begin{ruledtabular}
\begin{tabular}{|cc|ccc|} \
$\omega_r$ \ & \ $\omega_z$ \ & \ rescaling factor \ & \
${\mathcal{E}}_\Delta$ \ & \ $\lambda_\Delta^2$ \\
\hline \hline
$\hat{\omega}$ \ & \ $\hat{\omega}$
\ & \ $ $
\ & \ $ 1340 $ \ & \ $ 14.68 $ \ \\ \hline
\ $c_{\mathrm{cig}} \, \hat{\omega}$ \ & \
$c_{\mathrm{cig}} \, \hat{\omega}/5$
\ & \ $ c_{\mathrm{cig}} = 1.3463 $
\ & \ $ 1000 $ \ & \ $ 4.00 $ \ \\ \hline
\ $c_{\mathrm{pan}} \, \hat{\omega}/5$ \ & \
$c_{\mathrm{pan}} \, \hat{\omega}$
\ & \ $ c_{\mathrm{pan}} = 2.2447 $
\ & \ $ 550 $ \ & \ $ 1.05 $ \ \\
\end{tabular}
\end{ruledtabular}
\end{table}


\section{Lifetime of Condensates}
\label{Lifetime of Condensates}

In this section we first find expressions for the TIC, MQT and ICO
decay rates.
Using the numerical data presented in the previous section,
we then compute these decay rates for the
$\omega_r = \omega_z$ (isotropic),
$\omega_r /5 = \omega_z$ (cigar) and
$\omega_r = \omega_z / 5$ (pancake) cases.
Finally, we compare the results obtained for these
three potential geometries by studying the
spontaneous isotropization of the condensates.

\subsection{Definition and Computations of Decay Rates}
\label{Definition and Computations of Decay Rates}

The TIC (thermally induced collapse)
decay rate $\Gamma_T$ is estimated using the formula
\cite{Gardiner}
\begin{equation}
\label{eqT}
\frac{\Gamma_T }{ \hat{\omega}} =
\frac{ | \lambda_{+} | }{ 2 \pi }
\exp \left[ \frac{ - \hbar \hat{\omega}}{k_B T}
\left( {\cal{E}}_{+} - {\cal{E}}_{-} \right)
\right]
\end{equation}
%
where $\hbar \hat{\omega} ({\cal{E}}_{+} - {\cal{E}}_{-})$
is the (dimensionalized) height of the
nucleation energy barrier (with $\hat{\omega}$ the reference
frequency introduced in Section \ref{Gross-Pitaevskii Equation}), 
$T$ is the temperature of the condensate
and $k_B$ is the Boltzmann constant.
Note that the prefactor characterizes the typical decay time which is
controlled by the slowest part of the nucleation dynamics:
the top-of-the-barrier saddle point eigenvalue $\lambda_{+}$ and not
$\lambda_{-}$ as used in \cite{Stoof1}.
However, near the bifurcation both eigenvalues scale in the same way and
the behavior of $\Gamma_T$ can be obtained directly from
the universal saddle-node scaling laws (\ref{eqsclE})
and (\ref{eqscllam}). Thus the exponential factor and the prefactor
vanish respectively as $\delta^{3/2}$ and $\delta^{1/4}$.

\begin{figure}
\includegraphics[width=8cm]{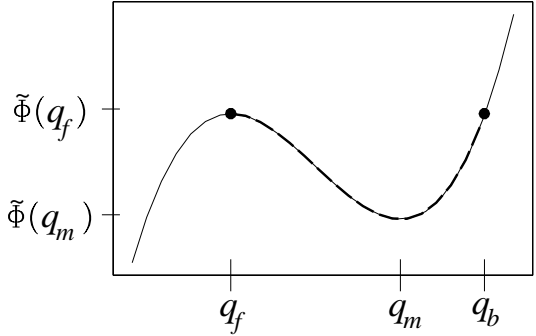}
\caption{\label{fig:potEff}
Bounce trajectory (dashed) over the Euclidean potential
$\tilde{\Phi}(q)$. Points $q_f$, $q_m$ and $q_b$ indicate the
fixed point, the minimum of $\tilde{\Phi}(q)$ and the bounce
point, respectively.
}
\end{figure}

We estimate the MQT (macroscopic quantum tunneling)
decay rate using an instanton technique
that takes into account the semi classical trajectory giving
the dominant contribution to the quantum action path integral
\cite{Stoof1,Ueda}.
We approximate this so-called bounce trajectory by the solution
of the equation of motion
\begin{equation}
\frac{d^2 q(t) }{ dt^2 } = \frac{- d\tilde{\Phi}(q)}{dq}
\end{equation}
starting and ending at the fixed point $q_f$ of the phase
space where ${\cal{E}}(q_f) = {\cal{E}}_{-}$.
The Euclidean potential $\tilde{\Phi}(q)$ is defined so that
$-\tilde{\Phi}(q)$ reconstructs the Hamiltonian
dynamics in the region scanned by the bounce trajectory
(see Figure~\ref{fig:potEff}).
 We represent it by a fourth-order polynomial of the form
\begin{equation}
\tilde{\Phi}(q)= \alpha_0 + \alpha_2 q^2 + \alpha_3 q^3 + \alpha_4 q^4
\end{equation}
coefficients $\alpha_0$, $\alpha_2$, $\alpha_3$ and
$\alpha_4$ chosen such that
\begin{subequations}
\begin{eqnarray}
\tilde{\Phi}(0) &=& -{\cal{E}}_{+} \\
\tilde{\Phi}(q_f) &=& -{\cal{E}}_{-} \\
\partial^2_{q} \tilde{\Phi}(0) &=& -\lambda_{+}({\cal{N}}) \\
\partial^2_{q} \tilde{\Phi}(q_f) &=& -\lambda_{-}({\cal{N}})
\end{eqnarray}
\end{subequations}
We thus obtain a semi-analytic polynomial expression
for $\tilde{\Phi}(q)$ where the coefficients are determined through the
numerical
values presented in figures \ref{fig:isotropic_eigs}, \ref{fig:cigar_eigs},
and \ref{fig:pancake_eigs}.

Once $\tilde{\Phi}(q)$ and the bounce point $q_b$
(defined through the relation $\tilde{\Phi}(q_b)=\tilde{\Phi}(q_f)$)
are known, the MQT rate is estimated as
\begin{equation}
\label{eqQ}
\frac{\Gamma_{Q}}{ \hat{\omega}} =
\sqrt{ \frac{|\lambda_{-}| v_0^2}{4 \pi} }
\exp \left[ \frac{-4}{\sqrt{2}} \int_{q_f}^{q_b}
\sqrt{\tilde{\Phi}(q)-\tilde{\Phi}(q_f)} dq \right],
\end{equation}
where $v_0$ is defined by the asymptotic form of the bounce trajectory
$q(t)$ as it approaches $q_f$ \cite{Stoof1}, given by
$q(\tau) \sim q_f + (v_0/|\lambda_{-}|) \exp[-|\lambda_{-} \tau|]$.

In the same way as was done for the TIC,
universal scaling laws can be derived close to criticality from
(\ref{eqNF}), (\ref{eqsclE}) and (\ref{eqscllam}).
We find that the exponential factor in (\ref{eqQ}) follows the same 
scaling as
$\sqrt{|{\cal{E}}_{+} - {\cal{E}}_{-}|} dq$. It therefore
vanishes as $\sqrt{\delta^{3/2}} \delta^{1/2} = \delta^{5/4}$.
The asymptotic form of $q(t)$ shows that 
$dq$ follows the same law as $v_0/|\lambda_{-}|$. Thus
$v_0 \sim \delta^{3/4}$, and
the prefactor vanishes as
$\sqrt{\delta^{1/4}} \delta^{3/4} = \delta^{7/8}$.
Note that these universal scaling laws agree with those already derived in
the Gaussian case in \cite{Ueda}.

\begin{figure}
\includegraphics[width=8cm,height=7cm]{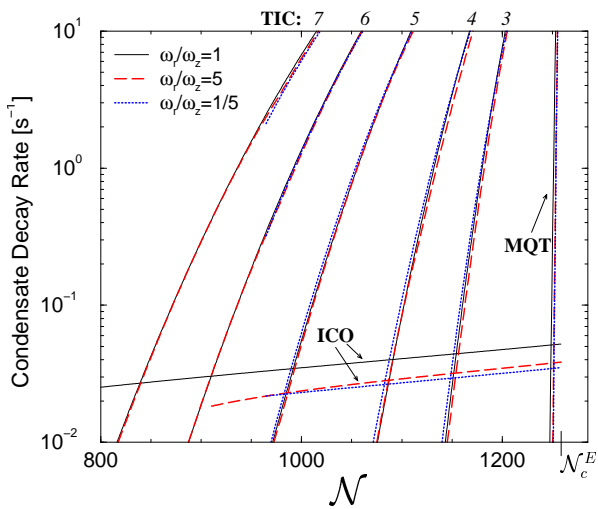}
\caption{\label{fig:decay_rates}
Condensate decay rates versus particle number for the isotropic
potential $\omega_r = \omega_z = \hat{\omega}$ (solid), and for
the rescaled cigar potential
$\omega_r = c_{\mathrm{cig}} \hat{\omega}$,
$\omega_z = c_{\mathrm{cig}} \hat{\omega} / 5 $
(dashed) and pancake potential
$\omega_r = c_{\mathrm{pan}} \hat{\omega}/5$,
$\omega_z = c_{\mathrm{pan}} \hat{\omega}$
(dotted).
ICO: inelastic collisions. MQT: macroscopic quantum tunneling.
TIC: thermally induced collapse at temperatures
$50$ nK (3), $100$ nK (4),
$200$ nK (5), $300$ nK (6), and $400$ nK (7).
}
\end{figure}

The TIC (\ref{eqT}) and MQT (\ref{eqQ}) decay rates
obtained for the exact and Gaussian stationary states
are shown in Figure \ref{fig:decay_rates}.
To validate these results we checked that the Gaussian TIC
decay rates computed in \cite{Sackett} are found
when we (incorrectly at a finite distance from criticality)
replace $\lambda_{+}$ by $\lambda_{-}$
in equation (\ref{eqT}) (data not shown).
We also checked that our Gaussian MQT decay rate agrees with the
one previously computed in \cite{Ueda}.

\begin{figure}
\includegraphics[width=8cm,height=7cm]{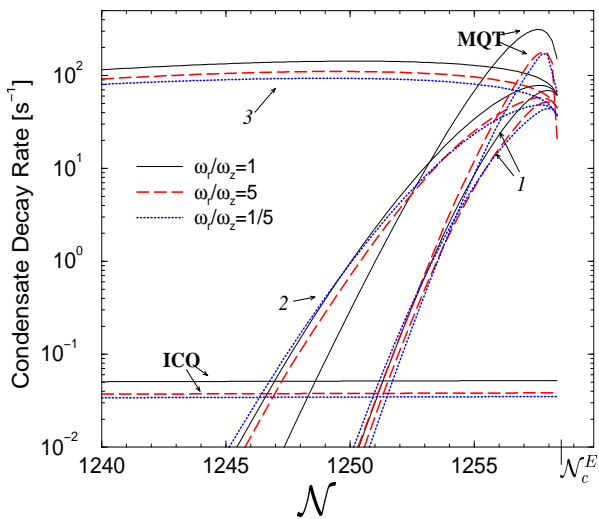}
\caption{\label{fig:decay_rates_zoom}
Enlargement of the crossover region between the
quantum tunneling and the thermal decay rate.
ICO: inelastic collisions. MQT: macroscopic quantum tunneling.
TIC: thermally induced collapse at temperatures
$1$ nK (1), $2$ nK (2) and $50$ nK (3).
}
\end{figure}

The ICO (inelastic two and three body collision)
atomic decay rates are evaluated using the formula
$d{\cal{N}}/{dt}=f_{C}({\cal{N}})$ with
\begin{equation}
\label{col}
f_{C}({\cal{N}}) =
K \int |\Psi|^4 d^3 {\bf x} + L \int |\Psi|^6 d^3 {\bf x},
\end{equation}
where
$K = 3.8 \times 10^{-4} [s^{-1}]$ and
$L = 2.6 \times 10^{-7} [s^{-1}]$
as in \cite{Shi,Dodd}.
In order to compare the particle decay rate (\ref{col})
to the condensate collective decay rates
obtained for the TIC and MQT, we compute the condensate ICO half-life using

\begin{equation}
\label{eqtau}
\tau_{1/2}({\cal{N}}) = \int_{{\cal{N}}/2}^{{\cal{N}}} \frac{dn}{f_C(n)}
\end{equation}
and plot $\tau_{1/2}^{-1}$ on figure \ref{fig:decay_rates}.

Figures \ref{fig:decay_rates} and \ref{fig:decay_rates_zoom}
compare the condensate decay rates for the
isotropic and the cigar and pancake potentials, rescaled by
$c_{\mathrm{cig}}$ and $c_{\mathrm{pan}}$ as described in Section
\ref{Non-isotropic Condensates}.
We note that the three aspect ratios generate very similar results
after rescaling.
The relative magnitudes of the different decay rates --
TIC, MQT, and ICO -- are the same for the three cases.
At $T \leq 1[\mathrm{nK}]$ the MQT effect becomes important compared to the
ICO decay in a region very close to ${\cal{N}}^E_c$
($\delta \leq 8 \times 10^{-3}$).
This was shown in \cite{Ueda} using Gaussian computations but
evaluating them with
the exact maximal number of condensed particles ${\cal{N}}^E_c$.
Figure \ref{fig:decay_rates_zoom} shows that even
for temperatures as low as $2 [\mathrm{nK}]$,
the TIC decay rate exceeds the MQT rate
except in a region extremely close to ${\cal{N}}_c$
($\delta < 5 \times 10^{-3}$),
where the condensates will live less than $10^{-1} [s]$.
Thus, in the experimental case of $^7$Li atoms, the
relevant effects are ICO and TIC, with the crossover
that is shown on Figure \ref{fig:decay_rates_zoom}.

\subsection{Spontaneous Isotropization of Condensates}
\label{Spontaneous Isotropization of Condensates}

\begin{figure}
\includegraphics[width=8cm,height=7cm]{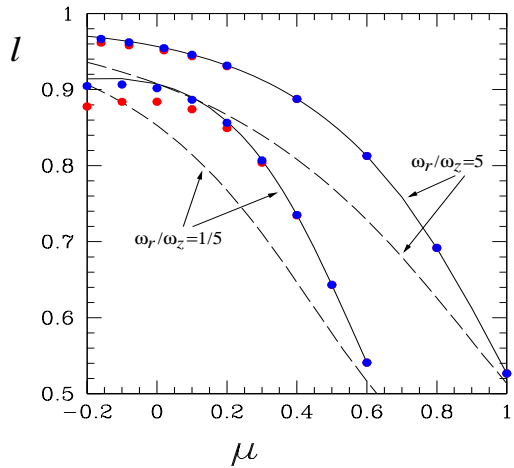}
\caption{\label{fig:iso}
Ellipticity ratio $\ell$ as a function of $\mu$.
Solid curves show numerical results, dashed curves
the Gaussian approximation.
For the cigar, $\ell \equiv \ell_r/\ell_z$ (upper curves) and
for the pancake, $\ell\equiv\ell_z/\ell_r$ (lower curves).
Dots show results obtained by using successively fewer Fourier 
modes in the numerical results; dots nearer to (further from) 
each curve correspond to retaining 7/8 (6/8) of the Fourier modes.
$\ell$ changes by less than 1\% for $\mu > -0.8$ ($\mu > -0.2$)
for the cigar (pancake) case and by less than 0.1\% at
the saddle-node bifurcation at $\mu = 0.38$ ($\mu = 0.31$) for
the cigar (pancake) case.
}
\end{figure}
\begin{figure}
\includegraphics[width=8cm,height=7cm]{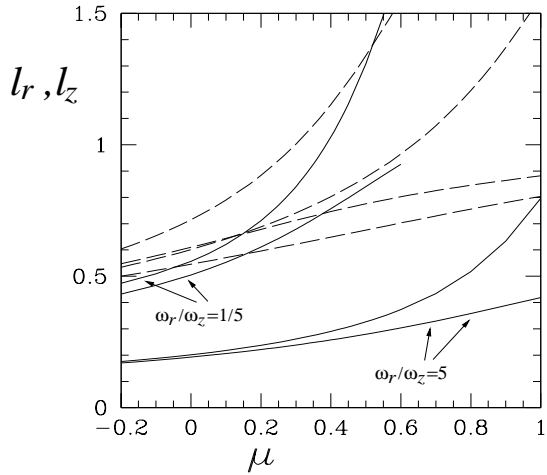}
\caption{\label{fig:size}
Length scales $\ell_r$, $\ell_z$ as a function of $\mu$.
Solid curves show numerical results, dashed curves
the Gaussian approximation.
Cigar case shown in lower curves (with $\ell_r < \ell_z$),
pancake in upper curves (with $\ell_z < \ell_r$).
The size of the condensate decreases drastically as
$\mu$ decreases, i.e. away from the saddle-node bifurcation
along the unstable branch.
}
\end{figure}
%
The decay rates of the isotropic and non-isotropic cases shown in
figure \ref{fig:decay_rates} are quite similar, despite the fact that
$\omega_z$ and $\omega_r$ differ by a factor of five. We have
investigated this question by examining the wave functions $\Psi$ for
the pancake and cigar cases. These wave functions are peaked at
the origin, as shown on figure \ref{fig:psi}.
Their characteristic length scales in the
axial and radial directions, $\ell_z$ and $\ell_r$, can be measured by
computing the ratios of the value of $\Psi$ to its curvature at the origin.

More specifically, we define
\begin{subequations}
\begin{eqnarray}
\ell_z^2 &\equiv& 
\Psi \left(\frac{\partial^2 \Psi}{\partial z^2}\right)^{-1} (r=0,z=0)\\
\ell_r^2 &\equiv& 
\Psi \left(\frac{1}{r} \frac{\partial}{\partial r} r 
\frac{\partial \Psi}{\partial r}\right)^{-1} (r=0,z=0)
\end{eqnarray}
\end{subequations}
We then obtain the ellipticity of the wavefunction as
the ratio $\ell$ of these length scales:
$\ell = \ell_r/\ell_z$ for the cigar and $\ell = \ell_z/\ell_r$ for the
pancake.
These ellipticity ratios are shown in figure \ref{fig:iso} as a function of
$\mu$.
For large $\mu$, i.e. away from the saddle-node bifurcation
along the stable branch, $\ell$ decreases rapidly away from one,
indicating that the wave function is highly non-isotropic.
At the saddle-node bifurcation, $\ell = 0.89$ at $\mu=0.38$ for the
cigar and $\ell=0.80$ at $\mu=0.31$ for the pancake.
As $\mu$ is decreased, i.e. as we leave the
saddle-node bifurcation along the unstable branch,
$\ell$ approaches one as the wave
function becomes more nearly spherically symmetric.
This trend is present both in
the numerical solution to the G-P equation and in the
Gaussian approximation, as can be seen on figure
\ref{fig:iso}. Since the decay rates result from the
scaling behavior near the saddle-node bifurcation, where the condensate
is fairly isotropic, it follows that the decay rates are similar
for the cigar, pancake, and spherically symmetric geometries, 
as we have shown
on Figures \ref{fig:decay_rates} and \ref{fig:decay_rates_zoom}.

The spontaneous isotropization of condensates when $\mu$ is decreased
can be understood by the following phenomenological reasoning. When
$-\mu$ grows, the balance of terms in the right hand side of equation
(\ref{NLS}) changes. For small $-\mu$, it is
dominated by the the isotropic $\nabla^2$ and the anisotropic $V(x)$
terms. But for large $-\mu$, the wavefunction $ \Psi$ is strongly peaked
and the
$\nabla^2$ and nonlinear terms, both isotropic, become dominant.

Figure \ref{fig:iso} also provides a test of our numerical
spatial resolution. By computing the ellipticity $\ell$
for different Fourier truncation levels, we show that
$\ell$ changes with the resolution for
low $\mu$, especially for the pancake case, where we used fewer
Fourier modes than in the other calculations.
Note however that our decay rate calculations only use results near the
saddle-node bifurcation, where $\ell$ varies
by less than 0.1\% when different truncation levels are used.
This indicates that $\Psi$ was adequately resolved in the region of
interest.

As $\mu$ is decreased, the wave functions become more highly peaked
for both our numerical results and for the Gaussian approximation
(see figures \ref{fig:psi} and \ref{fig:size}).
This is the main reason for the declining accuracy.
To continue the computations further,
the size of the periodic box should be reduced along with $\mu$.
We believe that, with adequate resolution, all of the exact
wave functions would become spherically symmetric
as $\mu$ decreases, as do the Gaussian approximations.

\section{Conclusion}
\label{Conclusion}
We have demonstrated that it is possible to numerically compute
the stationary states, the bifurcating eigenvalues and the
lifetime of anisotropic attractive Bose Einstein condensates.

The Gaussian mean field approximation was found to have
significant quantitative errors for all the different confining potential
geometries that were studied, when compared with numerical solutions
to the G-P equation.

Spontaneous isotropization of the metastable condensate
was found to occur as the critical number of particles is approached,
yielding a life-time that depends weakly on the anisotropy of the
confining potential.

Direct methods -- Gaussian elimination and diagonalization --
were used in treating the spherically symmetric case, of size $128$,
but are far too costly for the three-dimensional case, of size $10^6$.
In fact, since we only calculated axisymmetric
stationary states and eigenvectors with an additional midplane
symmetry, an intermediate two-dimensional axisymmetric cylindrical
representation could have been implemented, of size 5000, permitting
the use of direct methods.
Our purpose, however, has been to construct and explore numerical
methods appropriate for a general non-isotropic case.

The methods used to compute stationary states and bifurcating
eigenvalues for the non-isotropic cases where essentially analogous.
Each consists of a powerful and rapid
outer iteration: Newton's method for the stationary states
and the inverse Arnoldi method for the eigenvalues.
The large linear systems that need to be inverted within each method
are solved by the same inner biconjugate gradient iteration -- BiCGSTAB --
and constitutes the main numerical difficulty.
Its convergence is greatly improved by an inverse Laplacian
preconditioning which is empirically tuned by adjusting the
pseudo-timestep $\sigma$ in Newton's method or the shift $s$ in
Arnoldi's method.

Our results and implementation have demonstrated that all these
numerical techniques can be successfully combined
to calculate the stationary states and eigenvectors for
the G-P equation in a confining potential with
an arbitrary three-dimensional geometry.

\vskip 1cm
{\bf Acknowledgments}:
This work was supported by ECOS-CONICYT program no. C01E08
and by NSF grant DMR-0094569.
Computations were performed at the Institut du
D{\'e}veloppement et des Ressources en Informatique Scientifique
(IDRIS) of the CNRS.
\appendix*
\section{Numerical Methods}
\label{Numerical Methods}
\subsection{Spatial Discretization}
\label{Spatial Discretization}

The operators $L$ and $W$ defined in (\ref{Ldef}) and (\ref{Wdef})
are spatially discretized using the
pseudospectral method \cite{Got-Ors}. For the isotropic case,
the spherically symmetric $\Psi(r,t)$ is expanded as a series
of even Chebyshev polynomials $T_{2n}(r/R)$,
on which the boundary condition $\Psi(R,t) = 0$ is imposed.
The domain is taken to be $0\leq r\leq R=4$ and the resolution
used is $N_R=128$.
For the non-isotropic cases, we use a
three-dimensional periodic Cartesian domain and $\Psi$ is expanded
as a three-dimensional trigonometric (Fourier) series.
The cigar case is solved in a periodic domain of size
$(L_x,L_y,L_z)=(5.39, 5.39, 12.04)$ in units of $L_0$,
using $(N_x,N_y,N_z)=(96,96,96)$
gridpoints or trigonometric modes (with a
$2/3$ dealiasing rule,) so the total number of 
gridpoints or trigonometric functions is as high as $N_{3D}=10^6$.
(The more poorly resolved pancake case was calculated using
$(L_x,L_y,L_z)=(12.04, 12.04, 5.39)$ and
$(N_x,N_y,N_z)=(48,48,96)$.)
The harmonic potential (\ref{Vdef}) is approximated by
a periodic potential by writing $x=\arcsin(\sin(x))$
and Taylor-expanding the $\arcsin$ function. This leads
to a Fourier series for the potential, which is truncated
according to the resolution used.

Pseudospectral methods require performing over $\Psi$,
at every iteration, a Chebyshev transform in the isotropic case or a
Fourier transform in the non-isotropic case.
These operations  consume a time proportional to
$N_R \log N_R$ or $N_{3D} \log(N_{3D})$, respectively.
Actions and inversions of the Laplacian $L$ are carried out
on the Chebyshev or Fourier representations of $\Psi$, while
actions of the multiplicative operator $W$ are carried
out on its grid representations.
The time required by these operations scales approximately
linearly in $N_R$ or $N_{3D}$.

\subsection{Stationary States}
\label{Stationary States}

As stated in section \ref{Stationary}, the stationary states of (\ref{NLS})
that correspond to minima of ${\cal{E}}$ at a given value of ${\cal{N}}$
can be obtained by integrating to relaxation the diffusion equation
\begin{equation}
\label{GLR}
\frac{\partial \Psi}{\partial t} = L \Psi + W (\Psi)
\end{equation}
where the initial data $\Psi(t=0)$ has a total number of particles
${\cal{N}}$
and the value of the Lagrange multiplier $\mu $
is fixed during the relaxation by the condition
$ \partial {\cal{N}} / \partial t = 0$.

To integrate (\ref{GLR}) a mixed implicit-explicit first-order
time-stepping scheme is used:
\begin{equation}
\label{time_step}
\Psi(t+\sigma) = (I-\sigma L)^{-1} (I+\sigma W)\Psi(t)
\end{equation}
where $I$ is the identity operator.
The Helmholtz operator $(I-\sigma L)^{-1}$ is easily
inverted in the Chebyshev or Fourier representation.
The motivation for integrating $L$ implicitly is to
avoid the extremely small timesteps that would
otherwise be necessitated by the wide range
of eigenvalues of the Laplacian.

This relaxation method is equivalent to
that used in \cite{Ruprecht} and
can only reach the stable stationary solutions of (\ref{GLR}).
In order to also capture unstable stationary solutions \cite{Seydel}
we implemented a Newton branch-following algorithm \cite{PRLiso,TuckBark}.
We search for fixed points of (\ref{time_step}), 
a condition strictly equivalent to the stationarity of (\ref{NLS}): 
\begin{eqnarray} 
\label{fixed} 
0 = B\Psi(t) &\equiv& \Psi(t+\sigma) - \Psi(t) \nonumber\\ 
&=& (I-\sigma L)^{-1} (I+\sigma W)\Psi(t) - \Psi(t) \nonumber\\ 
&=& \left[(I-\sigma L)^{-1} (I+\sigma W) - I\right]\Psi(t) \nonumber\\ 
&=& (I-\sigma L)^{-1} \left[(I+\sigma W) - (I-\sigma 
L)\right]\Psi(t)\nonumber\\ 
&=& (I-\sigma L)^{-1} \left[(\sigma (L+W)\right]\Psi(t). \label{fixedpt} 
\end{eqnarray} 
Solutions to (\ref{fixedpt}) are found using Newton's method. 
We begin with an initial estimate $\Psi$, in our case the 
solution at a neighboring value of $\mu$. 
Newton's method calls for approximating the nonlinear operator $B$ 
whose roots are sought by its linearization $B_\Psi$ about $\Psi$.
We seek a decrement $\psi$ such that $\Psi - \psi$ 
solves this linearized equation: 
\begin{eqnarray} 
0 &=& B(\Psi - \psi) \approx B(\Psi) - B_\Psi\psi \nonumber \\ 
B_\Psi \psi &=& B(\Psi). \label{Newton1} 
\end{eqnarray}
$\Psi$ is then replaced by $\Psi-\psi$ and equation 
(\ref{Newton1}) solved again for a further decrement. 
The process is iterated until $B(\Psi)$ or $\psi$ 
is sufficiently small. 
In our case, equation (\ref{Newton1}) takes the form 
\begin{equation} 
(I-\sigma L)^{-1} \sigma (L+DW)\psi = 
(I-\sigma L)^{-1} \sigma (L+W)\Psi. \label{Newton2} 
\end{equation} 
We will explain how we solve the large linear problem (\ref{Newton2}) in 
section \ref{Conjugate Gradient Solution of Linear Systems}.

The role of $\sigma$ is formally that of the timestep 
in (\ref{time_step}), but in (\ref{Newton2}), 
its value can be taken to be arbitrarily large. 
For $\sigma \rightarrow \infty$, (\ref{Newton2}) becomes: 
\begin{equation} 
L^{-1} (L+DW)\psi =L^{-1} (L+W)\Psi. \label{Newtoninf}
\end{equation}

For the spherically symmetric case, the linear system
(\ref{Newtoninf}) is of size $N_R=128$
and can be solved by standard Gaussian elimination.
The boundary condition
$\psi(r=R)=0$ is imposed by modifying the operator
$L^{-1}$ or $(I-\sigma L)^{-1}$, as it is in the
time-stepping algorithm (\ref{time_step}).

To compute the full branch of solutions as a function
of $\mu$, we begin from a stable state of (\ref{GLR})
at a small value of ${\cal{N}}$ obtained by time-integration.
Each stationary state is computed in $3-5$ Newton
iterations.

\subsection{Conjugate Gradient Solution of Linear Systems}
\label{Conjugate Gradient Solution of Linear Systems}
For the periodic Cartesian case, the linear system
(\ref{Newtoninf}) of size $N_{3D}=10^6$
is too large to be stored or inverted directly: the
operation count for Gaussian elimination would be of the
order of $N_{3D}^3$.
Instead, we use BiCGSTAB \cite{Vandervorst}, a variant
of the well-known conjugate gradient method,
developed for linear systems which are not symmetric definite.
Such methods are matrix-free, meaning that they require
only the right-hand-side of (\ref{Newtoninf}),
and a subroutine which acts
with the linear operator of the left-hand-side.
A solution to the linear system is constructed as a
carefully chosen linear combination of powers of the
linear operator acting on the right-hand-side.

For a periodic Cartesian geometry, conjugate gradient
methods are particularly economical, since operator actions
are all accomplished in a time proportional to $N_{3D}$.
However, conjugate gradient methods for nonsymmetric definite
systems may converge slowly (requiring a large
number of evaluations of the linear operator) or even
not at all. This happens when the operator is poorly conditioned,
i.e. roughly when it has a wide range of eigenvalues.
One must then precondition the linear system,
i.e. multiply both sides of the system by a matrix which
improves its conditioning and accelerates convergence.
Since for operators such as $L+DW$, the wide range of
eigenvalues is due primarily to those of $L$, we
expect $L^{-1}$ to be an effective preconditioner.
From (\ref{Newton2}), it can be seen that $\sigma$
allows us to interpolate between linear operators
$\sigma(L+DW)$ and $L^{-1}(L+DW)$. We vary $\sigma$
empirically to optimize the convergence of BiCGSTAB.
A few hundred BiCGSTAB iterations are usually
required to solve the linear system.

A further advantage of iterative inversion methods
is that they can produce a (non-unique) solution
even when the linear operator is singular.
This is the case for our operators, which
have the neutral modes described in
section \ref{Linearized Stability}, as
well as other neutral modes related to
symmetries and the Fourier representation.
The preconditioner, however, is inverted
exactly. If (\ref{Newtoninf}) is used, the constant
Fourier mode is treated separately which allows us to
construct an invertible version of the singular operator $L$.

\subsection{Eigenvalue Problem}
\label{Eigenvalue Problem}

We now describe our numerical method for
calculating the linear stability of the stationary states.
For the spherically symmetric case,
the eigenvalues of (\ref{matrix})
are computed by constructing and
diagonalizing the corresponding matrix
for each converged stationary solution.
The results reported were generated
with a Mathematica code running on a workstation.
With the values $R=4$, $N_R=128$, the first two eigenvalues
of the harmonic oscillator are obtained with
a precision better than $0.05 \%$.

For the three-dimensional case, it is again not possible
to construct and diagonalize the matrix of size $N_{3D}=10^6$
directly: the operation count for diagonalization is also of
the order $N_{3D}^3$.
Instead, we calculate only eigenvalues of interest,
using a variant of the iterative inverse power method.
The inverse power method calculates the
eigenvalues of a matrix $M$ closest to a value $s$
by means of the sequence defined by:
\begin{equation}
(M - s I)\psi_{j+1} = \psi_j
\label{invpow}\end{equation}
The sequence $\{\psi_j\}$ converges rapidly to the
eigenvector whose eigenvalue is nearest to $s$,
with the eigenvalue $\lambda$ of $M$ estimated by
$(\lambda-s)^{-1} \approx
\langle\psi_{j+1},\psi_j\rangle/\langle\psi_j,\psi_j\rangle$.

In order to calculate complex or multiple eigenvalues,
and to obtain more precise eigenvalues and error estimates,
we use the sequence generated by (\ref{invpow}) to implement
the more general Arnoldi or Krylov method \cite{Arnoldi}.
Instead of retaining only the last two members of the sequence,
the last $K$ members (typically 4 or 6) are orthonormalized and then
assembled into the $K\times K$ matrix
$H_{jk}\equiv \langle \psi_j,(M-sI)^{-1}\psi_k\rangle$.
The eigenvalues of $H$ provide estimates of up to $K$
of the eigenvalues $(\lambda-s)^{-1}$ of $(M-sI)^{-1}$.

In our implementation of the Arnoldi method for (\ref{proba}),
we seek the eigenvalues $\lambda^2$ of the matrix
$-(L+DW^I)(L+DW^R)$.
Rather than solving (\ref{invpow}), we solve the equivalent
preconditioned problem \cite{TuckBark}
\begin{equation}
L^{-2}[-(L+DW^I)(L+DW^R) - sI]\psi_{j+1} = L^{-2}\psi_j
\label{precondeigprob}\end{equation}
by using BiCGSTAB.
From (\ref{precondeigprob}) we obtain a
sequence of vectors containing an increasing
proportion of the desired eigenvectors, but
since our solution of (\ref{precondeigprob}) is not exact, 
we then construct $H$ by multiplication
rather than inversion via
$H_{jk}\equiv \langle \psi_j,M\psi_k\rangle$.
We can then estimate the eigenvalues $\lambda^2$ by those of $H$.
Although the formal role of $s$ is that of a shift which
focuses the inverse iteration on the eigenvalues being
sought, here we also use it empirically to improve
the convergence of BiCGSTAB.

The inverse Arnoldi method requires between 3 and 10 iterations
to converge, each of which requires several hundred BiCGSTAB
iterations in order to solve its associated linear system.


\bibliographystyle{prsty}

\end{document}